\documentstyle[aps,epsf,twocolumn]{revtex}

\begin{document}
\title{A spin-coherent semiconductor photo-detector for quantum communication}
\author{Rutger Vrijen$^1$ and Eli Yablonovitch$^2$}
\maketitle \noindent \it{$^1$ Sun Microsystems Laboratories,
Mountain View, California\\ $^2$ University of California, Los
Angeles, Electrical Engineering Dept., Los Angeles, California\\ }
\newcommand{\degr}{$^{\circ}$}

\abstract{We describe how quantum information may be transferred
from photon polarization to electron spin in a semiconductor
device. The transfer of quantum information relies on selection
rules for optical transitions, such that two superposed photon
polarizations excite two superposed spin states. Entanglement of
the electron spin state with the spin state of the remaining hole
is prevented by using a single, non-degenerate initial valence
band. The degeneracy of the valence band is lifted by the
combination of strain and a static magnetic field. We give a
detailed description of a semiconductor structure that transfers
photon polarization to electron spin coherently, and allows
electron spins to be stored and to be made available for quantum
information processing. }

\section{Introduction}
\label{sec:introduction} \noindent The development of a quantum
computer will likely be preceded by the technology for
telecommunication of quantum information. The ability to transfer
quantum information from one physical form to another will become
valuable . One physical form, a photon in a coherent superposition
of polarization states, is particularly suitable for quantum
information transport, while another, for example electron spins
in a semiconductor host, or the internal states of atoms in a
high-Q cavity, may be preferred for quantum information storage
and processing. Transfer from one to the other will enable quantum
repeaters, quantum networks, and quantum entanglement over long
distances. In the short run, such technology, in the form of
quantum repeaters, will permit long-distance transmission of
quantum cryptographic keys over ordinary fiber.

Several proposals exist to transfer quantum information from
photons to atoms trapped in high-Q optical
cavities\cite{cirac-97,pellizari-97,vanEnk-97}. In this paper we
will describe how we may use opto-electronic technology to
transfer the quantum information stored in a coherent
superposition of photon polarization states to a coherent
superposition of electron spin states in a semiconductor. Such
electron spin states can have very long dephasing times, or $T_2$,
making them stable against qubit errors. In Silicon at low
temperatures, e.g. 1~K, measurements of $T_2$ indicate a lower
limit of 0.5~millisecond, limited by the isotopic purity of the
material. It is expected that in isotopically pure silicon, in
which all nuclei have spin zero, the $T_2$ will even be longer
since the most important decoherence mechanism, spin-diffusion
will be eliminated. Similarly long $T_2$ times are expected to
occur in low-temperature Germanium. Such long $T_2$ times allow
for extensive quantum information processing and error correction
or temporary storage of the information.

For coherent optical detection of quantum states however, we will
need a device with extremely high quantum efficiency, which can
only be achieved in the direct-bandgap III-V materials. In III-V
semiconductors, all nuclei have a net spin, and the electron $T_2$
times are significantly shorter. However, they are still long
enough, on the order of 100~nanoseconds\cite{awschalom-99}, that
quantum logic operations may be performed, particularly the
teleportation algorithm.

\section{Quantum State Coherent Photodetector}
\label{sec:detection} \noindent It is well known that polarized
electrons can be created by illuminating GaAs with circularly
polarized light. The valence band of GaAs has angular momentum
$L=1$. Combined with the spin of the electron $S=\frac{1}{2}$,
this gives rise to four $J=L+S=\frac{3}{2}$ states. The conduction
band wavefunction is an $S$-wave, with $J=\frac{1}{2}$. A
righthanded circularly ($m_j=1$) polarized photon, only couples
the valence state with $m_j=-\frac{3}{2}$ to the conduction band
$m_j=-\frac{1}{2}$ state. Thus, upon irradiation with righthanded
circularly polarized light, only electrons with spin down are
created. Similarly, left handed circularly polarized light only
creates spin up electrons. This polarization sensitivity is in
fact used in photocathodes for spin polarized high-energy electron
accelerators\cite{khateeb-99}.

However, this system is not suited for transferring quantum
information, encoded in an arbitrary superposition of polarization
states, from a photon to an electron. Consider a photon qubit, in
a superposition of circular polarization states:
\begin{equation}
\label{eq:photon}
\mid \phi \rangle_{ph}=\alpha \mid \sigma^+ \rangle + \beta \mid \sigma^- \rangle
\end{equation}
Since the two polarizations couple two different valence band states to the respective
conduction band states, they will also create two different hole states. The
electron thus created will be entangled with the hole, with the final state
of the electron-hole pair given by:
\begin{eqnarray}
\mid \psi \rangle_{eh} & = & \alpha \mid m_j=-\frac{3}{2}\rangle_h \otimes
    \mid m_j=-\frac{1}{2}\rangle_e + \nonumber \\
&  & + \beta \mid m_j=\frac{3}{2}\rangle_h \otimes \mid m_j=\frac{1}{2}\rangle_e
\label{eq:entangled}
\end{eqnarray}
This is an undesirable state, since in order to preserve the
quantum information, or process it, we would have to maintain
coherence for both the electron and the hole. Most likely the hole
will interact with the rest of the system and the superposition
will collapse into one of the two eigenstates in
Eqn~(\ref{eq:entangled}) of the system, and destroy the quantum
information.

To avoid this problem, we have to make sure that no information is
left behind in the hole state. Therefore, the two conduction band
polarizations have to be accessible from a single valence band
state. The two orthogonal photon polarizations should couple the
single valence state to both conduction band electron spin states.
Here we describe two important configurations in which this is
possible. Both rely on the capability to create a completely
non-degenerate topmost valence band state. This valence band can
be optically selected by using photons of an energy sufficient to
couple only that state to the conduction band. The other valence
bands are energetically too far removed from the conduction band
and can not be excited.

The degeneracy between heavy hole states ($m_j=\pm \frac{3}{2}$)
and light hole states ($m_j=\pm \frac{1}{2}$) is lifted if the
material is placed under strain, by growing heterostructures of
materials with different lattice constants, or by size quantization
in a quantum well. Generally, in
compressively strained materials, the heavy hole band is the
topmost band, while materials under significant tensile strain
have the light hole band on top. The growth direction $G$ thus
establishes a quantization axis. Finally, the remaining degeneracy
between the spin up and spin down components of both bands is
lifted by a static magnetic field. This magnetic field is chosen
small enough, so that the Zeeman perturbation is small compared to
the strain splitting.

\subsection{Case A: Magnetic Field Normal to Sample Surface}
\label{subsec:lighthole}
\begin{figure}[p]
\centerline{\epsfxsize=250pt
        \epsfbox{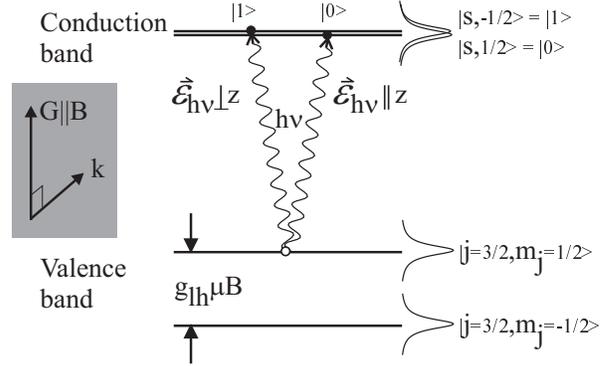}
} \vspace{-45ex} \caption{The excitation of a coherent
superposition of conduction band spin states, after absorption of
a photon in an identical superposition of polarization states. The
$k$-vector of the incident light is perpendicular to the magnetic
field, $\vec{k} \perp \vec{G} \parallel \vec{B}$. The two linear
polarization states of the photon are horizontal and vertical. The
spin states are energetically separated by a static magnetic field
$\vec{B}$ parallel to the growth direction $\vec{G}$. Horizontally
polarized photons have an electric field perpendicular to the
static magnetic field, and excite only $m_s=-\frac{1}{2}$
conduction band electrons. The electric field of vertically
polarized photons is parallel to the static magnetic field, and
excites only perpendicular $m_s=\frac{1}{2}$ electrons.}
\label{fig:lighthole}
\end{figure}

\noindent
In a sufficiently tensile strained semiconductor, the topmost valence band is the light hole band.
By applying a magnetic field $B$ parallel to the growth direction $G$ the band is split into its two Zeeman sublevels. The
energy splitting between these two states is given by
\begin{equation}
\label{eq:g}
E_z=g_{lh} \mu B
\end{equation}
where $g_{lh}$ is the Land\'{e} $g$-factor of the valence band.
The conduction band spin levels are also split by the magnetic
field, but the Zeeman energy can be very different from that of
the conduction band. The conduction band $g$-factor, $g_{cb}$ can
have a much smaller value than $g_{lh}$, as will be discussed
below. Accordingly, we can have the situation in which the
incident light has a bandwidth larger than the conduction band
Zeeman energy, but smaller than the valence band Zeeman energy.
Thus, we have a well defined, single initial valence-band state,
while the two conduction band spin states are practically
degenerate. The resulting level configuration is given in
Fig~(\ref{fig:lighthole}). The uppermost valence level, with
$m_j=\frac{1}{2}$, forms the single valence band state from which
both conduction band levels are accessed. We can see how this is
possible by expanding the $\mid J=\frac{3}{2}, m_j=\frac{1}{2}
\rangle$ in the LS basis:
\begin{equation}
\mid \psi \rangle_{lh}= \sqrt{\frac{2}{3}} \mid m_l=0, m_s=\frac{1}{2} \rangle +
\sqrt{\frac{1}{3}} \mid m_l=1, m_s=-\frac{1}{2} \rangle
\label{eq:wavefunction}
\end{equation}
The photon only interacts with the spatial part of the
wavefunction. Photons polarized along the quantization axis, i.e.
in the z-direction, do not change the $m_l$ number, and thus only
couple the $m_s=\frac{1}{2}$ part of the wavefunction to the
conduction band, which is exclusively $m_l=0$. Alternatively,
photons polarized perpendicular to the quantization axis change
the magnetic quantum number by 1, and thus exclusively interact
with the $m_s=-\frac{1}{2}$ part of the wavefunction, by coupling
the valence $m_l=1$ state to the conduction $m_l=0$ state.
/noindent Thus each polarization component exclusively excites one
electron spin component, while the hole state remains common to
both. Explicitly, if we have a photon qubit in an arbitrary
superposition of the two linear polarization states $\mid x
\rangle$ and $\mid z \rangle$:
\begin{equation}
\mid \phi \rangle_{ph} = \alpha \mid z \rangle + \beta \mid x
\rangle \label{eq:photqubit}
\end{equation}
it will excite an electron-hole pair with a wavefunction
\begin{eqnarray}
\mid \psi \rangle_{eh} & = & \alpha \mid j=\frac{3}{2},m_j=\frac{1}{2} \rangle_h \otimes
\mid l=0, m_s=\frac{1}{2} \rangle_e + \nonumber \\
 &  & + \beta \mid j=\frac{3}{2},m_j=\frac{1}{2}
\rangle_h \otimes \mid l=0, m_s=-\frac{1}{2} \rangle_e \nonumber \\
 & = & \mid j=\frac{3}{2},m_j=\frac{1}{2} \rangle_h \otimes
 \nonumber \\
 & & \otimes  \{ \alpha \mid l=0,m_s=
 \frac{1}{2} \rangle_e + \beta \mid l=0,m_s=-\frac{1}{2} \rangle_e \}
\label{eq:superpos}
\end{eqnarray}

As is obvious from Eqn~(\ref{eq:superpos}), the state of the hole
can be factored out of the wavefunction of the system and no
information is carried away by the hole. The electron and hole are
disentangled. Furthermore, the quantum information originally
stored in the photon, is completely transferred to the electron
spin.

\subsection{Case B: Magnetic Field Parallel to Sample Surface}
\noindent In Part A, the case of magnetic field normal to the sample
surface was treated. As shown in Eqn.~\ref{eq:wavefunction}, the weighting
amplitude for optical absorption into up versus down spin differs by $\sqrt{2}$.
This can be compensated by using an off-center wavelength for the optical
transition, but it can also be eliminated by magnetic field parallel to the
sample surface as follows. With magnetic field parallel to the sample surface, a
new quantization axis is created, in which the regular quantum eigenstates of
the normal axis case are equally super-posed.

For case A, the heavy hole valence band was not useful, since the $m_j=+3/2$ heavy
hole states coupled purely to $m_j=+1/2$ electron states, and were unable to create the required quantum mechanical superposition. With the magnetic field parallel to the
surface, a new quantization axis is created, that creates an equal superposition
but suffers from another problem. The component of the heavy hole $g$-tensor parallel
to the surface vanishes for heavy holes. Thus the requirement for an optically resolvable
valence band Zeeman splitting cannot be achieved for heavy holes. Now in case B, as previously in case A, heavy holes are not useful for creating superpositions of
quantum states. However, an equal superposition can be created in case B for a light hole valence band.  In this case the band will be split into two
levels $\mid \psi^+ \rangle$ and $\mid \psi^- \rangle$:
\begin{eqnarray}
\mid \psi^+ \rangle & = & \frac{1}{\sqrt{2}} \{
    \mid m_j=-\frac{1}{2} \rangle + \mid m_j=\frac{1}{2} \}  \nonumber \\
\mid \psi^- \rangle & = & \frac{1}{\sqrt{2}} \{
    \mid m_j=-\frac{1}{2} \rangle - \mid m_j=\frac{1}{2} \}
\label{eq:psi}
\end{eqnarray}
Using $\mid \psi^+ \rangle$ as the
initial state, it is possible to couple to the $m_s=-\frac{1}{2}$
conduction band state by exciting a conduction band electron with
a righthanded circularly polarized photon ($m_l=1$). At the same
time, a lefthanded circularly polarized photon will excite the
$m_s=\frac{1}{2}$ conduction band state. Again, the conduction
band $g_{cb}$-factor is much smaller than the valence band
$g_{lh}$-factor, so that both conduction band spin states are
energetically accessible. Note that the quantization axis for the
photons, or the $k$-vector, coincides with the strain direction in
the semiconductor, and is perpendicular to the static magnetic
field $\vec{B}$. Again, the non-degenerate valence band state is
the same for both cases, and no quantum information is retained in
the hole. One major difference with respect to case A
is that the excited conduction band electron state is not an
eigenstate of the system. This is due to the magnetic field being
perpendicular to the quantization axis. Therefore the spin
eigenstates of the conduction band electron are:
\begin{eqnarray}
\mid 0 \rangle = \sqrt{\frac{1}{2}} \{ \mid m_s=-\frac{1}{2} \rangle - \mid m_s=\frac{1}{2} \rangle \} \nonumber \\
\mid 1 \rangle = \sqrt{\frac{1}{2}} \{ \mid m_s=-\frac{1}{2} \rangle + \mid m_s=\frac{1}{2} \rangle \}
\label{eq:0}
\end{eqnarray}
After excitation of a pure $ m_s=-\frac{1}{2} $ or a pure $
m_s=\frac{1}{2} $ the conduction band electron spin will precess
around this magnetic field. The precession time is given by
$\tau=2 \pi\hbar / g_{cb} \mu B$. Let us consider the excitation
of the system with a photon carrying quantum information:
\begin{equation}
\mid \phi \rangle_{ph} =\alpha \mid \sigma^+ \rangle + \beta \mid \sigma^- \rangle
\label{eq:photqubit2}
\end{equation}
The system, initially in $\mid \psi^+ \rangle$ is excited to a superposition:
\begin{eqnarray}
\mid \psi \rangle_{eh} & = & \alpha \mid \psi^+ \rangle_h \otimes \mid m_s=-\frac{1}{2}
\rangle_e +  \beta \mid \psi^+ \rangle_h \otimes \mid m_s=\frac{1}{2} \rangle_e \nonumber \\
& = & \mid \psi^+ \rangle_h \otimes \{ \frac{\alpha}{\sqrt{2}}(\mid 0 \rangle
+ \mid 1 \rangle) + \frac{\beta}{\sqrt{2}}(\mid 0 \rangle - \mid 1 \rangle) \}_e
\end{eqnarray}
\begin{figure}[p]
\centerline{\epsfxsize=250pt
        \epsfbox{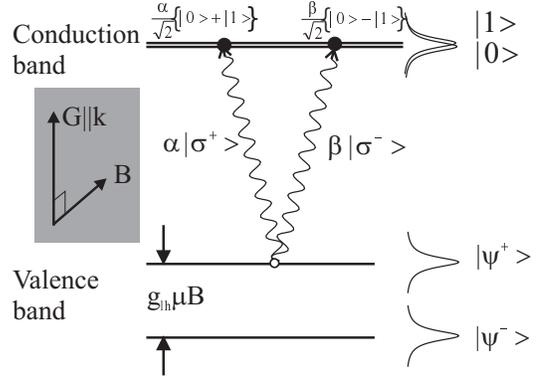}
} \vspace{-45ex} \caption{The excitation scheme for magnetic field
parallel to the sample surface. The $k$-vector of the circularly
polarized incident light is parallel to the growth direction $G$
and perpendicular to the magnetic field, $\vec{k}
\parallel \vec{G} \perp \vec{B}$. Left hand circularly polarized($m_l=-1$)
photons interact only with the $m_l=1,m_s=\frac{1}{2}$ component
of the initial state $\mid \psi^+ \rangle$, exciting only  $\mid
m_s=\frac{1}{2} \rangle =\sqrt{\frac{1}{2}} \{ \mid 0 \rangle +
\mid 1 \rangle \}$ conduction band electrons. Similarly, the right
hand circularly polarized photons only excite $\mid
m_s=-\frac{1}{2} \rangle$.} \label{fig:heavyhole}
\end{figure}
\noindent As long as the coherence is preserved, i.e. within a
$T_2$, the precession of the electron spin does not destroy the
quantum information present in the superposition of the two spin
states. A well-synchronized Hadamard transform will take the
system from a superposition of precessing states to a
corresponding superposition of the two eigenstates $\mid 0
\rangle$ and $\mid 1 \rangle $. The quantum information has again
been successfully transferred to the spin of the electron, without
any entanglement with the hole that is left behind.

\subsection{g-factors of valence-  and conduction band}
\noindent Essential to the proposed detection mechanisms is that
both Zeeman levels of the conduction band electron are accessible
within the bandwidth of the exciting photon, while the two Zeeman
levels in the valence band are spectrally resolved. Therefore it
is crucial to find, or rather engineer, a material in which the
g-factor of the conduction band electron is significantly smaller
than the g-factor of the valence band hole. Fortunately, in III-V
semiconductors, the valence band states have rather large
spin-orbit coupling which allows engineering of the g-factor. An
example is the InAs/GaAs quantum well as described in
ref\cite{sirenko-96}. In this system, the conduction band electron
g-factor is $g_{cb}=0.4$, while the light-hole $g$-factor can as
large as $g_{lh}=8.87$. Such a large difference in $g$-factor
allows the valence bands to be energetically resolved, while, at
the same magnetic field, the conduction band spin states are
spectrally overlapping.

\section{Spin Coherent Photo Detector and Emitter}
\subsection{Detector}
\label{subsec:detector} \noindent
\begin{figure}[p]
\centerline{\epsfxsize=250pt
        \epsfbox{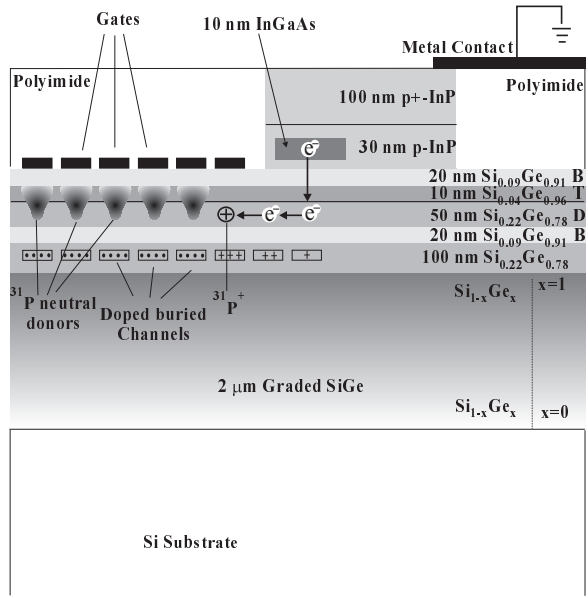}
} \vspace{-25ex} \caption{A conceptual side view cross section of
the spin-coherent photo-detector. The detector contains an
optically active 10-nm InGaAs quantum well embedded in p-doped
InP, where the photon excites an electron. The quantum information
is transferred from the photon polarization to the electron spin
in this step. The III-V material is wafer fused to a SiGe
heterostructure. After the electron is generated it is transported
to the SiGe material. There it is bound to a positive $^{31}$P
donor ion. The electron remains at the donor location, but the
quantum information can be swapped further into the information
processor by spin-exchange operations, controlled by the electric
gates located a above the $^{31}$P neutral dopants.}
\label{fig:receiver}
\end{figure}
\noindent In Fig~(\ref{fig:receiver}) a schematic illustration of
the proposed spin coherent receiver is given. In the device we
combine silicon and III-V technology. The materials are joined by
the technique of wafer fusion. This is a commercial process, and
it allows us to combine the high optical efficiency of III-V
semiconductors with the extremely long $T_2$ lifetimes that
electron spins enjoy in a type IV semiconductor host, particularly
silicon. The electron is generated in a 10-nm thick InGaAs quantum
well, which is embedded in p-doped InP. The quantum well is
optically thin, so that the absorption needs to be enhanced by an
optically resonant cavity\cite{campbell-99}.

After generation, the hole is swept to ground through a heavily
doped $p^+$-InP layer. The hole carries no quantum information
whatsoever, i.e. is not entangled with the spin states of the
conduction band electron. The electron is transported through the
wafer-fused layer into the storage and processing part of the
spin-coherent receiver by electrostatic fields. Such transport can
be performed while retaining coherence to a remarkably high
degree, as was recently demonstrated in an experiment by Awschalom
et al\cite{awschalom-99}. The storage and processing part of the
detector is in fact a small quantum processor. This indicates how
the first application of small quantum computers may in fact be to
boost the throughput of quantum communication channels. The logic
gates are in the SiGe heterostructure, which has been described in
detail elsewhere\cite{vrijen-99}. The layers with a
Si$_{0.09}$Ge$_{0.91}$ composition, labeled B, are barrier layers
with a relatively high bandgap, to confine the electron in the
region of interest. Between these layers, the electron is bound to
a $^{31}$P$^+$ ion, which is implanted underneath an electric
gate. There the electron remains bound, but the information stored
in its spin states can be swapped further into the computer,
through spin-exchange, or two-qubit, operations with neighboring
electrons bound to identical $^{31}$P dopant ions, also placed
under gates. The required control is obtained by applying gate
voltages, and by the fact that the $g$-factor of the bound
electron is modulated between the two barrier (B) layers. In the
Ge-rich Si$_{0.04}$Ge$_{0.96}$-layer, also called T for tuning
layer, the $g$-factor is Ge-like, and has a value of $g=1.563$ in
the $\langle 100 \rangle$ direction . In the Si-rich
Si$_{0.22}$Ge$_{0.78}$, or D(onor), layer, the $g$-factor is
Si-like, with $g=1.998$. By modulating the gate voltage, the
electron wave-function is pulled more or less toward the electrode
through the Stark-effect. This in turn modulates the $g$-factor.
In the presence of a constant background microwave radiation
field, the electron spin can thus be pulled in and out of
resonance. This allows single-qubit interactions. The two-qubit
interaction is achieved through the application of positive gate
voltages to adjacent electrodes. Both electrons are pulled away
from their ions, and the reduced Coulomb interaction allows an
increased Bohr radius. This causes an increased interaction
between the neighboring electrons, and turns on the exchange
interaction. As shown by Loss et al\cite{loss-98}, this exchange
interaction can be used to perform swap, and controlled NOT
operations between qubits. These one- and two-qubit interactions
together form a universal set of quantum gates, so that any
unitary transformation can be performed on the quantum information
stored in the bound electron spins, including error correction and
quantum information processing.

\subsection{Photon Emitter}
\label{subsec:emitter} 
\begin{figure}[p]
\centerline{\epsfxsize=250pt
        \epsfbox{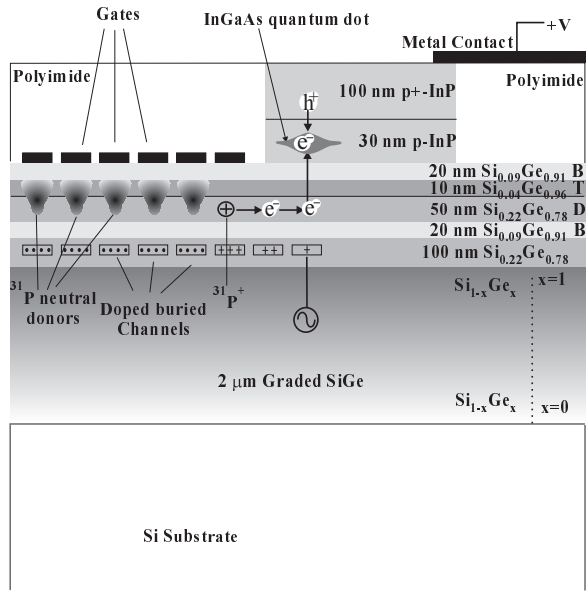}
} \vspace{-25ex} \caption{A cross section of the spin-coherent
emitter. The structure is very similar to the receiver, but there
are a few crucial differences. After processing or storage of the
quantum information, it is swapped back to the rightmost $^{31}$P
donor. This donor is ionized with an electric field pulse, and the
electron is transported into a quantum dot in the optically active
III-V material. The quantum dot provides a high degree of
confinement in all dimensions, so that the electron is not free to
move. This is important since the III-V environment contains a lot
of nuclear spins, and the electron would sample a lot of
inhomogeneities in the local nuclear spins if it was free to move
around, which could lead to rapid decoherence. With the electron
in the quantum dot, a hole is injected, and a photon is emitted
upon recombination of the electron and hole. The same selection
rules that govern the transfer of information in the detection
step, provide the reverse transfer in the emission step.}
\label{fig:emitter}
\end{figure}

\noindent After error correction,
teleportation, or other quantum information processing steps, we
want to re-emit the stored quantum information as a photon. This
is possible by running the reverse of the detection process. The
electron needs to be recombined with a hole, where only one type
of hole is allowed to interact with the electron. Since we chose
our holes to be in the top of the valence band, this will not be a
problem since they will naturally be the abundant ones. The photon
emitter therefore looks very similar to the photo-receiver, as
shown in Fig~(\ref{fig:emitter}).

The information that needs to be transmitted is first swapped to
the electron bound to the rightmost $^{31}$P dopant ion. This
electron is pulled off the ion by applying a relatively strong
pulse to  the gate electrode, and is electrostatically transported
to the InGaAs quantum dot. A hole is injected into the dot as well
so that electron-hole recombination can take place. Upon
recombination the quantum information is transferred to the
emitted photon, following the same selection rules that governed
the transfer in absorption.
\noindent The quantum dot is a slight modification with respect to
the quantum well in the receiver and ensures that the electron
wavefunction is confined in space and is not free to move around
in the time before a photon is emitted. In the receiver, the
electron is swept away from the III-V material as soon as it is
generated. In the emitter however, we have to wait for the
electron to interact with the single injected hole, which may take
several nanoseconds. A free electron would sample many different
spins in such a lifetime, and decohere rapidly. By confining the
electron to a dot, such movement is precluded. The quantum dot
needs only to be small enough so that the confinement energy of
the quantum dot is larger than $kT$, and so that the charging
energy satisfies $e/C > kT$ as well\cite{yamamoto-99}.
Furthermore, to assure a long enough capture time, the tunneling
resistance $R$ for electrons flowing into the quantum dot should
be $R>26 k \Omega \approx h/e^2$. All these requirements can be
met simultaneously even with a moderate sized quantum dot,
patterned by e-beam lithography.

One complication in the emitter is that, whereas the incident
photon in the detector had a definite $k$-vector, the emitted
photon, can be emitted isotropically. Not all directions obey the
same selection rules however, since the relative orientation of
the spin-quantization axis and the photon $k$-vector varies with
the emission direction. Therefore it may be necessary to introduce
complex quarter-wave plates or phase plates to compensate for the
photon polarization differences in different directions in the
optical collection system.

Summarizing, we have indicated how optical selection rules may be
used to transfer quantum information from one physical form to
another, i.e. from photon polarization to electron spin and vice
versa. A key ingredient is the removal of all degeneracies from
the initial valence band state, to prevent entanglement of the
conduction band electron spin state and the remaining hole. We
have described a detector and emitter device combining a optically
highly efficient III-V component with a type-IV storage and
processing component, in which the electron spins have exceedingly
long $T_2$ lifetimes, based on such selection rules. Such a device
may become an indispensable component in quantum communication
technology by boosting the data rate of existing channels, and by
enabling the long-distance transfer of fully quantum states.

The authors would like to acknowledge fruitful discussions with
Tal Mor, Oscar Boykin, Vwani Roychowdhury and Ivair Gontijo. This
work was sponsored by the Defense Advanced Research Project Agency
(Grant No. DAAD19-00-1-0172) . The content of this paper does not
necessarily reflect the position or policy of the Government, and
no official endorsement should be inferred.


\end{document}